# Electrical property of Half Metallic Ferromagnet $Pr_{0.95}Mn_{0.939}O_3$

B. Santhosh Kumar, P. Manimuthu, N. Praveen Shanker and C.Venkateswaran*

*Department of Nuclear Physics, University of Madras, Guindy campus, Chennai – 600 025, India.*

*\*Corresponding author E mail: cvunom@hotmail.com*

**Abstract:**

Half metallic ferromagnetic compounds (HMF) are intensively explored for their potential in spintronics application. According to theoretical studies, HMF are those materials where the conduction mechanism is due to one type of spin state, either up or down. Band theory strongly predicts that Praseodymium (Pr) based transition metal compound (TMC) is HMF whose electrical properties are a consequence of the contribution from both the *4f* and *3d* orbitals of Pr and TMC ions. Hence, an investigation has been made by preparing $Pr_{0.95}Mn_{0.939}O_3$ by ball milling assisted sintering process. The phase confirmation of the prepared compound is carried out using X-Ray diffraction technique. Impedance analysis with respect to temperature indicates a decreasing trend in electrical conductivity with respect to temperature, confirming the semiconductor behaviour of the sample. This may be due to the consequence of the interaction of *4f* and *3d* orbitals of the rare-earth (Pr) and transition metal compound (Manganese).

**Introduction:**

The structural family of pervoskite forms a large family of compounds having crystal structure related to the $ABO_3$ form. The ideal pervoskite crystallizes to cubic $ABO_3$ pervoskite and can be described as consisting of corner sharing $[BO_6]$ octahedra with the A cation occupying the 12- fold coordination site formed in the middle of the cube of eight such octahedra [1]. The interest in compounds belonging to this family of crystal structure arises in large because of the surprising magnetic and electrical properties in single compound of pervoskite materials. These class of compounds provide the probe to control the magnetic property via electrical property and vice versa (ferroic order). These type of compounds offer wide range of application in relaxors, memory devices, capacitors and offer tunnelling property for switching purposes in some optical devices.

Among pervoskite structure, praseodymium (Pr) based manganese (Mn) oxides have attracted much attention in last few years due to the theoretical prediction of Pr and Mn based oxides as HMF. In HMF, the conduction mechanism is governed by either of the spin carriers (spin up or spin down) [2]. One spin state offers insulating property while the other offers metallic property. According to experiments, Pr based compounds offer interesting electrical and magnetic properties. Due to Mn, there will be spatially degenerate electronic ground state which undergoes some structural distortion which is called as Jahn teller (JT) distortion which gives interesting electrical and magnetic property suitable for device applications [3].

In $PrMnO_3$ (PO), both Pr and Mn exists in trivalent oxidation state. Therefore, one can expect a novel transport property with respect to temperature in PO due to JT. The Goldschmidt tolerance factor [1] allows us to estimate the degree of distortion. For PO, the tolerance factor is calculated as 0.886. PO crystalizes into orthorhombic structure (consistence with tolerance factor) with the space group of *pbnm*.

The variation of resistivity with respect to temperature is one of the notable phenomenon of any material. In $PrMnO_3$, one spin state act as a conductor while the other spin state act as an insulator. The switching over between two extreme (conducting and non-conducting) conditions with respect to temperature makes this type of material fit into many applications like switching, capacitance etc. Impedance spectroscopy plays a vital role in understanding the electrical transport properties of the material. In this technique, the sample is subjected to an alternating frequency range between 1 Hz to 10 MHz under various temperatures.

The purpose of this communication is to synthesis Pr based Mn oxide using high energy ball milling method and to investigate the electrical transport property of the polycrystalline sample.

**Experimental Detail:**

High quality precursor of $Pr_6O_{11}$ and $MnO_2$ were taken in a stoichiometric ratio in zirconia vial with a ball-to-powder ratio of 27:4 and subjected to planetary ball milling at 450 rpm for 5 hours. The collected sample was subjected to heat treatment at different temperatures and the best result was obtained by heating the sample at 1473 K for 10 hours and by slowly cooling it down to room temperature. The structural characterization of the sample was carried out using X-ray diffractometer (GE-XRD 3003 TT) with Cu Kα radiation (λ = 1.5406 Å). The electrical properties were investigated using an impedance analyzer in the temperature range of 303 K to 503 K (Solatron SI 1260 impedance/gain phase analyzer).

**Results and discussion:**

The collected sample is subjected to room temperature X-ray diffraction using Cu-K$\alpha$ radiation ($\lambda$ = 1.540598 Å) between the range of 20° to 70° for phase confirmation. Figure 1, confirms the formation of $Pr_{0.95}Mn_{0.939}O_3$ (PMO) with orthorhombic crystal system and space group of *pbnm* with one secondary peak due to formation of $PrMn_2O_5$ (in figure 1. patterned as *). It is also observed that the peaks of PMO merges together and there is a lateral shift in the pattern which is due to strain caused by ball milling preparation. The presence of secondary peak confirms that the prepared sample contains Mn in mixed valence state of $Mn^{4+}$ and $Mn^{3+}$. The secondary peaks may be due to lack of oxygen at high temperatures (At higher temp. the oxygen content inside the furnace decreases). Due to the different oxidation state of Mn ions, there is possibility of magnetic exchange between these ions which is termed as double exchange interaction.

Table 1: shows the lattice parameters of PMO and $PrMnO_3$. By comparing the lattice parameters of PMO (calculated and literature), a slight contraction along the *b* direction and elongation towards *a* and *c* direction is observed. This is due to the presence of $PrMn_2O_5$, which consists of $Mn^{4+}$ and $Mn^{3+}$ ions. Theoretically, it is reported that bond length between *Mn-O* is altered by $Mn^{3+}$ JT ion. $Mn^{4+}$ is JT inactive ion, whereas $Mn^{3+}$ is JT active [3].

*Table1: Shows the lattice parameters for PMO (calculated and literature) and $PrMnO_3$*

|  | a (Å) | b (Å) | c (Å) |
|---|---|---|---|
| PMO (calculated) | 5.465(1) | 5.498(5) | 7.740(5) |
| PMO (literature) | 5.4562 | 5.5914 | 7.672 |
| $PrMnO_3$ | 5.545 | 5.787 | 7.575 |

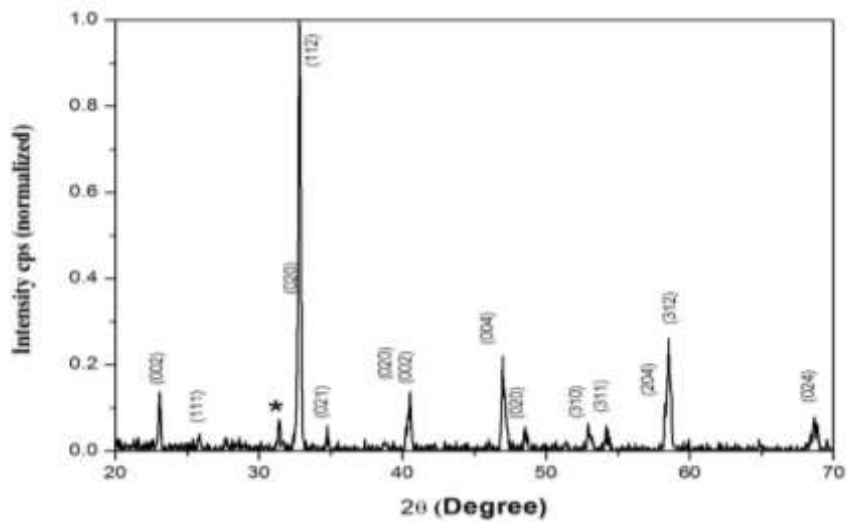

*Figure 1: XRD pattern of $Pr_{0.95}Mn_{0.939}O_3$ sample. The * (asterisk) symbol represent the secondary phase ($PrMn_2O_5$)*

The impedance spectroscopy technique has been carried out to analyze the electrical response of the prepared sample for a frequency range from 1 Hz to 10 MHz at different temperature from 303 K to 503 K. The sample was made into pellet form by applying a pressure of 2.5 ton.

From the impedance measurement, the resistivity vs temperature plot [5] is obtained as shown in figure 4. The figure shows the semiconducting (as the temp increases resistance decreases) nature of the sample with respect to temperature. According to bandtheroy prediction, there is an overlap of the orbitals from Pr based Mn oxides. So at higher temperatures, the overlapping of the *4f* and *3d* orbitals should be much closer, which decreases the resistivity of the sample with respect to temperature. As reported earlier [3], even the contribution from $Mn^{3+}$ and $Mn^{4+}$ ions decreases the resistivity of the sample. $Mn^{3+}$ Oxidises to $Mn^{4+}$ by introducing holes in 3d band which gives rise to a decrease in the value of resistivity of the sample. At this occasion, Pr exhibits a strong basic character, which is suitable to stabilize $Mn^{4+}$ with accurate structural parameters.

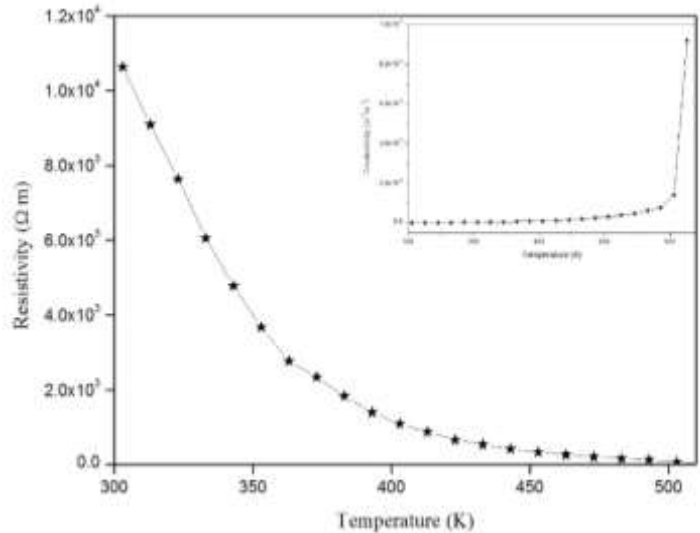

*Figure 2: Temperature dependent resistivity plot, insert figure show the conductivity plot Vs temperature*

**Conclusion:**

$Pr_{0.95}$ $Mn_{0.939}O_3$ was successfully prepared by ball milling assisted sintering method with the presence of secondary phase of $PrMn_2O_5$, as evident from XRD. The decreasing trend of temperature dependent resistivity plot confirms the overlap of *3d* and *4f* orbitals as predicted by Half Metallic Ferromagnetic theory (band theory). This kind of property will be suitable for switching devices and spintronics application.

**Acknowledgements**

The author B. Santhosh kumar (BSK) thank DST – Inspire for its financial support in the form of fellowship and also Mr B. Soundararajan, technical officer for his kind help and support.

**References:**

1. M.A.Pena and J.L.G.Fierro, Chem Rev.101, 1981 (2001).

2. J.M.D. Coey and M. Venkatesan, J.App. Phys.  91, 8345 (2002).

3. J.A.Alonso, M.J.Martinez and M.T.Casais, Inorg. Chem, 39, 917 (2000).

4. Sergio Ferro, Internation Journal of Electrochemistry 561204 (2011).

5. Dinesh Varshney, poova sharma and I.mansuri, AIP Conf. Proc. 1349, 941 (2011).